\documentclass[twocolumn,prb,aps,showpacs,superscriptaddress]{revtex4-1}
\pdfoutput=1
\usepackage{siunitx}
\usepackage{graphicx}
\usepackage{amsmath}
\usepackage{float}

\begin{document}

\title{Tailoring non-collinear magnetism by misfit dislocation lines}

\author{Aurore Finco}
\author{Pin-Jui Hsu}
\author{Andr\'{e} Kubetzka}
\author{Kirsten von Bergmann}
\author{Roland Wiesendanger}

\affiliation{Department of Physics, University of Hamburg, D-20355 Hamburg,
Germany}

\date{\today}

\begin{abstract}

The large epitaxial stress induced by the misfit between a triple atomic layer 
Fe film and an Ir(111) substrate is relieved by the formation of a dense 
dislocation line network. Spin-polarized scanning tunneling microscopy (SP-STM) 
investigations show that the strain is locally varying within the Fe film and
that this variation affects the magnetic state of the system. Two types of 
dislocation line regions can be distinguished and both exhibit spin spirals
with strain-dependent periods (ranging from \SI{3}{\nano\meter} to 
\SI{10}{\nano\meter}). Using a simple micromagnetic model, we attribute the 
changes of the period of the spin spirals to variations of the effective
exchange coupling in the magnetic film. This assumption is supported by the
observed dependence of the saturation magnetic field on the spin spiral period.
Moreover, magnetic skyrmions appear in an external magnetic field only in one
type of dislocation line area, which we impute to the different pinning
properties of the dislocation lines. 

\end{abstract}


\maketitle

\section{Introduction}

Strain-induced control of complex magnetic states such as spin spirals or
skyrmions~\cite{bergmann_interface-induced_2014, nagaosa_topological_2013} is a
multifaceted approach towards manipulation of spin structures. Mechanical or
piezoelectrical~\cite{cui_method_2013} setups as well as direct influence on
the sample growth are used to investigate such effects with multiple
experimental techniques on various systems. The observed phenomena are ascribed
to changes of the effective anisotropy: uniaxial mechanical compression tunes
the stability of the skyrmion lattice phase, e.g. in the helimagnet 
MnSi~\cite{nii_uniaxial_2015}, where hydrostatic pressure also decreases the
helix period and the critical temperature~\cite{fak_pressure_2005}.  On the
other side, epitaxial strain created by growing multiferroic BiFeO$_3$ thin
films on different substrates can destroy the cycloidal spin spiral and
stabilize an antiferromagnetic state~\cite{sando_crafting_2013}, again via
effective anisotropy. The non-collinear magnetic structures mentioned above are 
stabilized by the Dzyaloshinskii-Moriya interaction (DMI), which can also be 
affected by strain. This mechanism was found in FeGe, where uniaxial tensile 
strain dramatically distorts the skyrmion lattice~\cite{shibata_large_2015}.

Although previous studies were mostly focused on the effects of spatially 
uniform strain, we investigate here the influence of strain variations on a 
local scale in an ultrathin magnetic film using spin-polarized scanning 
tunneling microscopy (SP-STM)~\cite{wiesendanger_spin_2009}. Our system 
consists of three atomic layers of Fe deposited on Ir(111). Because of the
misfit between Ir (fcc, lattice constant \SI{3.84}{\angstrom}) and Fe (bcc,
lattice constant \SI{2.86}{\angstrom}), the triple layer Fe film exhibits a 
dislocation line network. We observe that the epitaxial strain relief is not
uniform within the ultrathin film, resulting in different regions exhibiting 
spin spirals with various periods and vanishing at different magnetic field 
intensities. We attribute these differences to spatial variations in the 
strength of the effective exchange coupling. Depending on the atom arrangement
in the magnetic film, these spirals may or may not split up into single
magnetic skyrmions~\cite{hsu_electric_2016}. 

\section{Experimental details}

The experiments were performed in a ultrahigh vacuum system with a base
pressure below $10^{-11}$ \si{\milli\bar}. Different chambers are used for
substrate cleaning, Fe deposition and STM measurements.  The Ir single crystal
substrate was prepared by cycles of Ar-ion sputtering at
\SI{800}{\electronvolt} and annealing up to \SI{1600}{\kelvin} for
\SI{60}{\second}. The Fe film was then evaporated onto the clean substrate at
elevated temperature (about \SI{200}{\celsius}) at a deposition rate around 0.2
atomic layer per minute.
Four different scanning tunneling microscopes were used for this work, three
low temperature microscopes with He bath cryostats operating at respectively
\SI{4}{\kelvin}, \SI{5}{\kelvin} and \SI{8}{\kelvin} as well as a variable
temperature system equipped with a He flow cryostat. Out-of-plane external
magnetic fields were applied during the low temperature measurements using
superconducting coil magnets. We used antiferromagnetic bulk Cr tips for the
spin-resolved measurements.
All the STM topography images were measured in constant-current mode, where
the stabilization current is kept constant by a closed feedback loop.
The differential conductance maps were simultaneously recorded at a fixed
sample bias voltage using lock-in technique.
	
\section{Morphology of the Fe film}

\begin{figure}
	\includegraphics[scale=1]{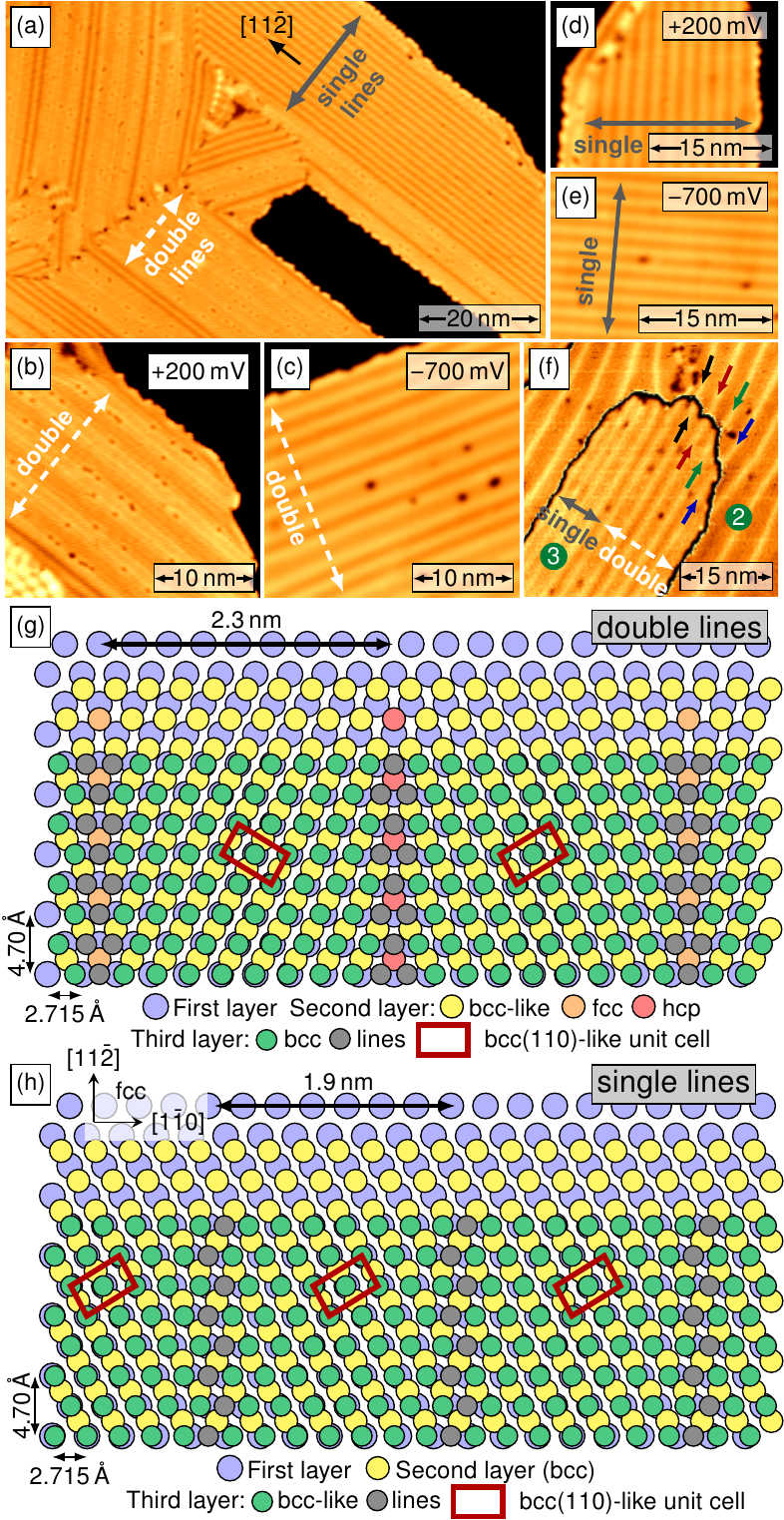}
	\caption{\label{structure} (a) Constant-current STM topography map of the
triple layer Fe film on Ir(111). Double and single lines are indicated by the
arrows. (b),(c) Zoom-in on double line areas. At positive bias, one can see
that bright and dark lines alternate and show a double line feature, whereas
at negative bias, the lines look all very similar. (d),(e) STM topography of
single line regions. The single lines have the same appearance at any sample
bias voltage. (f) Constant-current STM topography map of a triple layer Fe
island on top of a double layer Fe film. The numbers in green circles
indicate the local thickness of the film. The color scale was adjusted
separately for the two terraces in order to highlight the matching of the
dislocation lines. The areas between the bright lines on the double layer are
bcc. (g),(h) Proposed structure models for the triple layer Fe film from the
experimental observations presented in (a)-(f).
\emph{Measurement parameters:}  ${I = \SI{1}{\nano\ampere}}$ and
(a)~${U = +\SI{200}{\milli\volt}}$, ${T = \SI{8}{\kelvin}}$, 
${B = \SI{3.5}{\tesla}}$ ;
(b)~${T = \SI{5}{\kelvin}}$, ${B = \SI{4.5}{\tesla}}$ ;
(c)~${T = \SI{5}{\kelvin}}$, ${B = \SI{3}{\tesla}}$ ;
(d)~${T = \SI{8}{\kelvin}}$, ${B = \SI{0}{\tesla}}$ ;
(e)~${T = \SI{8}{\kelvin}}$, ${B = \SI{2.5}{\tesla}}$ ;
(f)~${U = \SI{-700}{\milli\volt}}$, ${T = \SI{153}{\kelvin}}$, 
${B = \SI{3}{\tesla}}$.}
\end{figure}

The STM topography map presented in Fig.~\ref{structure}(a) shows a typical
triple layer Fe film on an Ir(111) substrate. The surface of the film exhibits
a dense network of dislocation lines. These lines follow the three equivalent
high-symmetry directions $\left\langle 11\bar{2} \right\rangle$ of the fcc(111)
surface~\cite{hsu_electric_2016} as already reported for the double layer Fe
film~\cite{hsu_guiding_2016}.

Two types of dislocation lines can be distinguished from the STM image of
Fig.~\ref{structure}(a): at positive sample bias voltage, the dislocation lines
in the region spanned by the white dashed arrow show an alternating bright and
dark contrast as well as double line features (see detail in
Fig.~\ref{structure}(b)). These areas will thus be named \emph{double lines}
in the following. The spacing for double lines is ranging from
\SI{2.3}{\nano\meter} to \SI{3}{\nano\meter}, which corresponds to half
the structural period. On the other hand, in the region marked by the gray
arrow, the lines are denser (spacing between \SI{1.8}{\nano\meter} and
\SI{2.2}{\nano\meter}) and they all have the same appearance at any bias
voltage (see Fig.~\ref{structure}(d) and Fig.~\ref{structure}(e)). Hence
these lines will be designated as \emph{single lines}. At negative sample
bias voltage, the distinction between single and double lines from the sole
topography becomes challenging as illustrated in Fig.~\ref{structure}(c)
and Fig.~\ref{structure}(e).

Figures~\ref{structure}(g) and~\ref{structure}(h) show proposed atomic
structure models for the triple layer Fe film. In both cases, the first layer
Fe is pseudomorphic with respect to the Ir(111) substrate
lattice~\cite{heinze_spontaneous_2011}. The double lines are located exactly
on top of the dislocation lines of the double layer Fe film as shown in
Fig.~\ref{structure}(f). The double layer lines seem to get closer
when they approach the island: the strain relief is larger in the triple than
in the double layer. In the atomic model, the second layer is uniaxially
compressed with respect to a pseudomorphic layer and the arrangement of the Fe
atoms is getting close to bcc(110). The lines correspond to the positions where
the atoms are located on the fcc or hcp sites and the triple layer follows a
bcc stacking on top of the double layer. In the case of the single lines, also
the second layer grows pseudomorphically but in bcc stacking (see
Fig.~\ref{structure}(h)) and only the third layer on top is uniaxially
compressed. All the lines are identical and induced by lateral shifts of
the atoms. As will be shown later by looking at the magnetic structure, the
double line regions have mirror planes along the dislocation lines, whereas
two mirror symmetric domains exist for the single line regions,
in agreement with the structure model. 

\section{Zero field non-collinear magnetic structure}

\begin{figure}[H]
	\centering
	\includegraphics[scale=1]{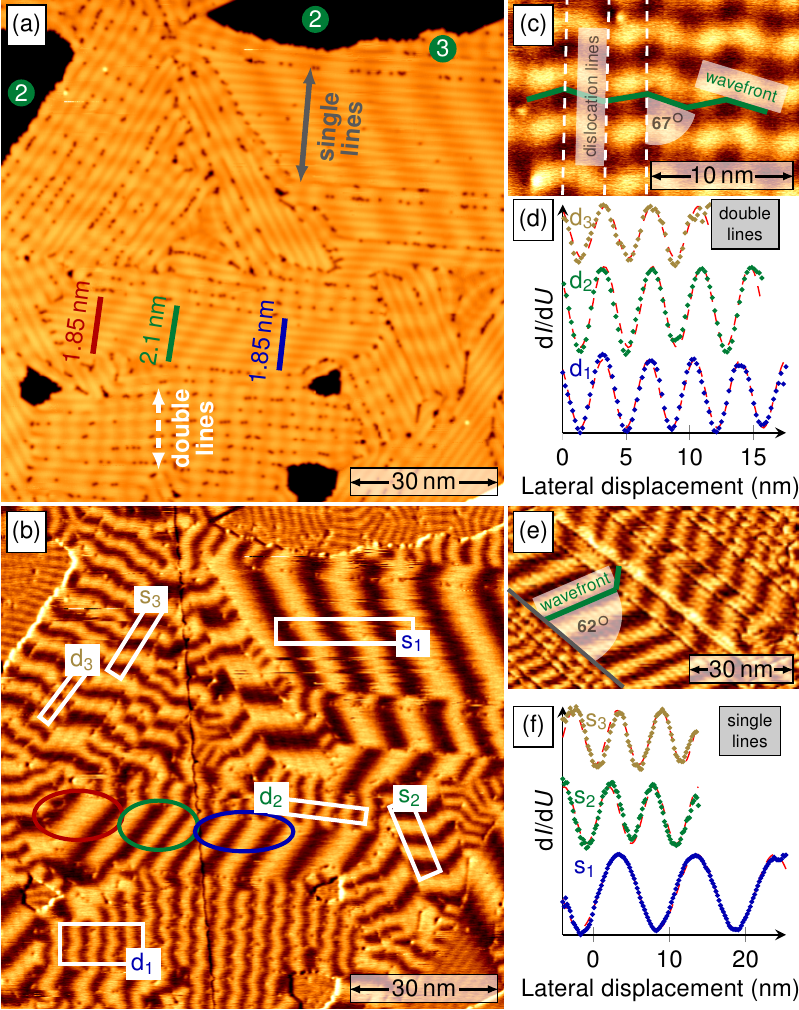}
	\caption{\label{magnetism} (a),(b) Spin-resolved constant-current map and
spin-resolved differential tunneling conductance map of a triple layer Fe film
on Ir(111) measured simultaneously using a Cr bulk tip with out-of-plane spin
sensitivity. The magnetic sensitivity of the tip is simply deduced from the
magnetic contrast on the spirals, which is identical for all the propagation
directions. The magnetic contrast is visible on both the constant-current and
differential conductance maps. The ellipses on image (b) mark a zone where the
period of the spiral is changing: this can be correlated with the spacing
between the dislocation lines indicated in (a). (c) Detail of the spin-resolved
differential tunneling conductance map of a double line region. The spin spiral
propagates along the lines and its wavefront exhibits a zigzag shape.
(d),(f) Line profiles taken in the double and single line regions,
respectively, at the positions indicated in (b). The displayed profiles are the
mean values within the rectangles, each data line was laterally shifted to
straighten the wavefront and offsets were applied to the resulting profiles for
clarity. The red dashed lines are the results of fits with sine functions,
showing that the spirals are homogeneous regardless of the considered region.
(e) Spin-resolved differential conductance map of an area with single lines.
Spin spirals with various wavelengths are visible and their wavefronts are
tilted by about $\pm \SI{62}{\degree}$ with respect to the lines.
\emph{Measurement~parameters:} no external magnetic field,
(a),(b)~${U = \SI{-700}{\milli\volt}}$, ${I = \SI{1}{\nano\ampere}}$,
${T = \SI{8}{\kelvin}}$ ;
(c)~${U = \SI{-700}{\milli\volt}}$, ${I = \SI{750}{\pico\ampere}}$,
${T = \SI{4}{\kelvin}}$;
(e)~${U = \SI{-700}{\milli\volt}}$, ${I = \SI{1}{\nano\ampere}}$, $
{T = \SI{8}{\kelvin}}$.}
\end{figure}

The difference between single and double line areas becomes more striking when
we look at the magnetic structure. Figures~\ref{magnetism}(a) and
\ref{magnetism}(b) show respectively a spin-resolved constant-current map and a
simultaneously spin-resolved differential conductance map measured with an
out-of-plane spin-sensitive Cr bulk tip at low temperature. These SP-STM
measurements reveal that spin spirals propagate along both types of dislocation
lines. The nature of the spin spirals is cycloidal: this has been shown already
for the double layer Fe spirals~\cite{hsu_guiding_2016} as well as for the
double line regions in the triple layer Fe~\cite{hsu_electric_2016} and is
assumed also to be the case for the single line areas since the Fe-Ir
interface-induced DMI is large~\cite{heinze_spontaneous_2011}. However, the
appearance of the spin spirals varies in the different regions of the Fe film.
The wavelength of the spirals is smaller in the double line regions
(\SI{3}{\nano\meter} to \SI{4.5}{\nano\meter}) than in the single line regions
(between \SI{5}{\nano\meter} and \SI{10}{\nano\meter}). 

Furthermore, the wavefront of the double line spirals has a zigzag
shape~\cite{hsu_guiding_2016, hsu_electric_2016}, whereas that of the spirals
in the single line regions is straight but tilted with respect to the lines as
can be seen from the details of spin-resolved differential conductance maps in
Fig.~\ref{magnetism}(c) and Fig.~\ref{magnetism}(e).
The proposed structure models can explain these shapes for the wavefronts:
the wavevector prefers to follow the bcc[001]-like rows of atoms as observed
for the double layer Fe on Ir(111)~\cite{hsu_guiding_2016},
Cu(111)~\cite{phark_reduced-dimensionality-induced_2014} and
W(110)~\cite{meckler_real-space_2009}. 
For the double lines, the direction of
the rows alternate and this creates the magnetic zigzag structure, whereas
the direction does not change for the single lines, as shown by the
bcc(110)-like unit cells marked in red in Fig.~\ref{structure}(g) and 
Fig.~\ref{structure}(h). However, for the single line regions, two mirror
symmetric structural domains are possible and found in the SP-STM data.

For a strict coupling of the wavevector to the bcc [001] direction, the
expected angle $\alpha$ between the wavefront and the dislocation lines
(as defined in Fig.~\ref{magnetism}(c) and Fig.~\ref{magnetism}(e)) can be
computed from the structure models:
	\begin{equation}
	 \tan \alpha = \frac{\sqrt{3} \delta}{\delta+a} 
	\end{equation}
	 where $\delta$ is the line spacing and ${a = \SI{2.715}{\angstrom}}$ the
in-plane interatomic distance for Ir(111). This expression is the same for both
structures. This angle is below \SI{60}{\degree} and decreases with the line
spacing. Yet the angle obtained from the data is slightly larger (up to
\SI{10}{\degree}) and seems to change randomly. This deviation which is going
towards a straighter wavefront for the double line regions was previously
observed for the double layer spirals~\cite{hsu_guiding_2016}. The guiding of
the wavevector might compete here with the presence of energetically
disadvantageous kinks in the wavefront. For the single line areas, the angle
is also larger, meaning that the wavefront is more perpendicular to the
dislocation lines than expected. This effect could be attributed to boundary
effects as well as to domain wall-like structures preferring to be as short
as possible. Line profiles of spirals (Fig.~\ref{magnetism}(d) and
Fig.~\ref{magnetism}(f)) taken in various regions can be fitted with sine
functions. This sinusoidal shape indicates that the spirals are homogeneous,
i.e. no direction of the magnetic moments is more favorable than the others.
Yet epitaxial strain in ultrathin films creates an effective anisotropy via
magnetoelastic  coupling~\cite{bruno_magnetic_1989} and since the strain
relief is not uniform in the triple layer Fe film, the total effective
magnetic anisotropy is expected to vary between the different regions. This
should result in more or less pronounced distortions of the spiral profiles
depending on the dislocation line spacing. We therefore conclude from the
homogeneity of all the observed spiral profiles that the effective magnetic
anisotropy and its variations in the triple layer Fe film are small enough to
be neglected in the following.

\section{Tuning of the spiral period with strain relief}

In Fig.~\ref{magnetism}(b), a single line area in which the wavelength of the
spiral is changing is marked with ellipses. This variation occurs because the
dislocation lines become more distant in the green zone than in the surrounding
red and blue ones. The distance between the lines in each ellipse is given in
the constant-current map of Fig.~\ref{magnetism}(a). The closer the lines, i.e.
the larger the compression in the third layer Fe, the larger the spiral period.

\begin{figure}
	\includegraphics[scale=1]{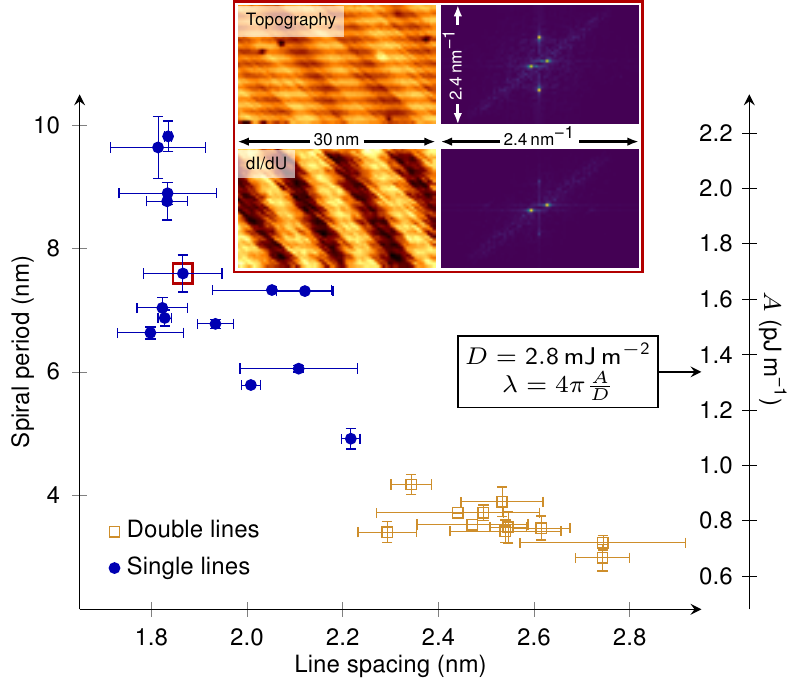}
	\caption{\label{plot_mag_strain} Dependence of the spiral period on the
spacing of the dislocation lines for the triple layer Fe on Ir(111), both
double and single line regions. The SP-STM measurements were performed at low
temperature (\SI{4}{\kelvin} or \SI{8}{\kelvin}) on several different samples
and the spiral periods were extracted using 2D Fourier transformation and
fits of the real space data to sine functions, keeping only points with an
error bar below 15\%. For the double line regions, the zigzag shape of the
wavefront was not considered and the period was measured along the
lines. The actual wavelength along the direction of the wavevector might be
10\% smaller. On the single line regions, the angle of the wavefront was
taken into account in the determination of the period.
It is evident that if the distance between the dislocation lines
decreases, the spiral period increases, as observed already in
Fig.~\protect\ref{magnetism}. The insets show the data (both spin-resolved
constant-current map and differential conductance map as well as their
Fourier transforms) used to determine the point marked with the red square.
The axis on the right indicates that the effective exchange parameter should
vary between \SI{0.6}{\pico\joule\per\meter} and
\SI{2.2}{\pico\joule\per\meter} to create spirals with wavelengths between
\SI{3}{\nano\meter} and \SI{10}{\nano\meter} from the simple model described in
the text.
}
\end{figure}	

The graph in Fig.~\ref{plot_mag_strain} presents a systematic investigation
of the effect of the strain relief on the wavelength of the spirals, for both
double and single line regions. The spiral period and the spacing between
dislocation lines  were measured using Fourier transforms and fits of the real
space data to sine functions in several different regions on several samples.
The trend observed in Fig.~\ref{magnetism} is confirmed, the spiral period
decreases when the line spacing increases. The line spacing is linked to the
amplitude of the strain relief by the structure models: the mean value of the
distance between the Fe atoms in the fcc [1\=10] direction in the top layer
decreases with the line spacing. To give an order of magnitude, this distance
is 5\% smaller for a \SI{1.8}{\nano\meter} than for a \SI{3}{\nano\meter}
line spacing.

When the distance between the dislocation lines changes, all the interatomic
distances in the Fe film are modified, thus it is expected that the strengths
of the magnetic interactions will be affected. The relevant interactions for
this system are the exchange couplings and the interface-induced
DMI~\cite{fert_role_1980, bergmann_interface-induced_2014}. Based on the data
shown in Fig.~\ref{magnetism}(d) and Fig.~\ref{magnetism}(f), we assume that
the effective anisotropy is negligible. The DMI originates mainly from the
Fe-Ir interface and since the first layer Fe layer is pseudomorphic regardless
of the considered region, it should not be significantly influenced (contrary
to bulk systems like FeGe~\cite{shibata_large_2015}). We therefore infer that
the strain relief is acting mostly on the exchange couplings to modify the
period of the spirals.

\textit{Ab initio} calculations for a free-standing Fe bcc(110)
layer~\cite{shimada_ab_2012} also assigned the period variation of the
spirals as well as their stability under in-plane strain to modifications of
the exchange couplings. However, the spin spiral state is only stable in the
free layer when a compressive strain is applied and the compression reduces
the period of the spiral. The calculated evolution of the spirals with the
strain is thus opposite to what we measured but this discrepancy could result
from the presence of the substrate and the complicated atomic structure of
the Fe film. 

In order to estimate the magnitude of the strain effect, we consider a
simplified one-dimensional micromagnetic model derived from the one proposed by
Bogdanov and Hubert~\cite{bogdanov_thermodynamically_1994} in which only the
effective isotropic exchange coupling and the DMI are kept. We completely
ignore here that the film is not spatially uniform, resulting in spatially
inhomogeneous magnetic coupling constants. Only average values are taken into
account. The energy density is thus:
\begin{equation}
	\label{energy_density}
	\mathcal{E} = A \sum_i \left(\frac{\partial \mathrm{\mathbf{m}}}{\partial x_i}
	             \right)^2 +
	             D \left( \mathrm{m}_z \frac{\partial \mathrm{m}_x}{\partial x} 
                 - \mathrm{m}_x \frac{\partial \mathrm{m}_z}{\partial x} \right)
\end{equation}
where $A$ is the effective exchange stiffness constant, $D$ the effective DMI
constant and $\mathrm{\mathbf{m}}$ the reduced dimensionless magnetization. In
this particular case, there is always a stable cycloidal spin spiral state in
the system and its period $\lambda$ is simply:
\begin{equation}
	\label{period}
	\lambda = 4\pi\frac{A}{D}
\end{equation}
For the monolayer Fe on Ir(111), the value of the DMI constant from
\textit{ab initio} calculations is ${|d| = \SI{1.8}{\milli\electronvolt}}$~
\cite{heinze_spontaneous_2011} which gives the micromagnetic parameter
${D = \SI{2.8}{\milli\joule\per\square\meter}}$ for a three-layer-thick Fe film.
As shown in Fig.~\ref{plot_mag_strain}, the exchange stiffness parameter
should then vary between \SI{0.6}{\pico\joule\per\meter} and
\SI{2.2}{\pico\joule\per\meter} to obtain magnetic periods ranging from
\SI{3}{\nano\meter} to \SI{10}{\nano\meter}. These values are similar to
${A = \SI{2.0}{\pico\joule\per\meter}}$ found for the PdFe bilayer on Ir(111)~
\cite{romming_field-dependent_2015} with a spiral period of about
\SI{6}{\nano\meter}. Since the DMI originates from the
interface, its effect should decrease when the thickness of the film increases,
i.e. the typical length scale of the magnetic structure is expected to become
larger. Indeed, the period of the nanoskyrmion lattice on the monolayer Fe on
Ir(111) is \SI{1}{\nano\meter}~\cite{heinze_spontaneous_2011}, the wavelength
of the double layer spiral is about
\SI{1.6}{\nano\meter}~\cite{hsu_guiding_2016} and for patches of a size below 
\SI{100}{\nano\meter} (larger ones were not found on the samples), the
quadruple layer appears ferromagnetic (as, for example, in
Fig.~\ref{hysteresis}).

\section{Skyrmions and domain walls in magnetic field}

\begin{figure}
	\includegraphics[scale=1]{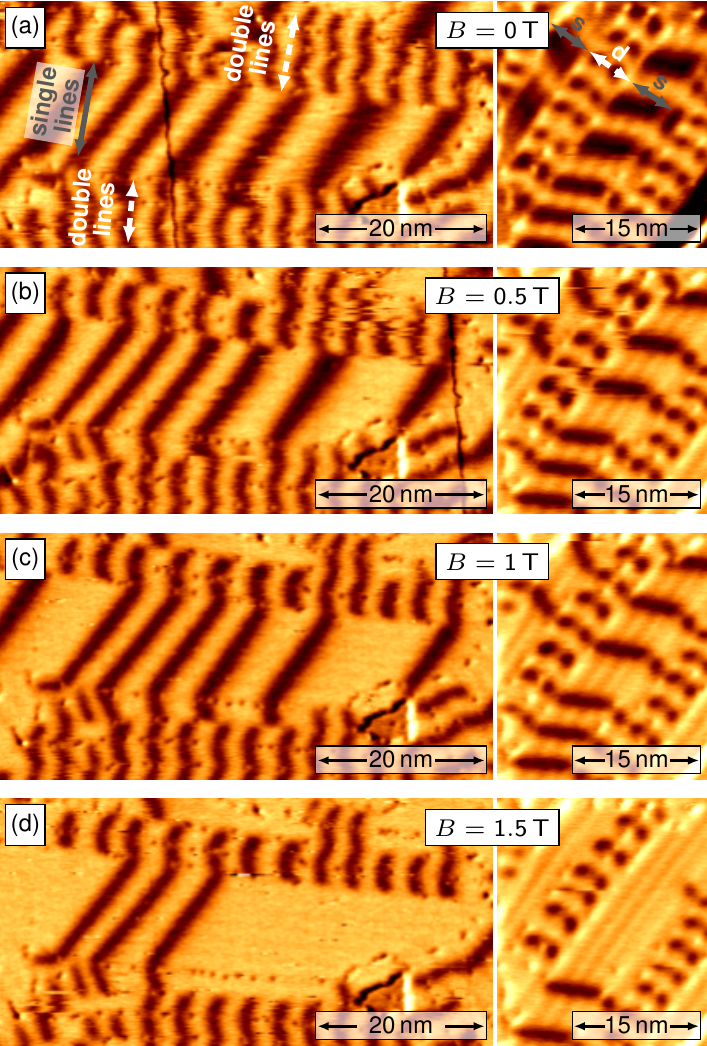}
	\caption{\label{adj_regions} Spin-resolved differential tunneling conductance
maps of a region with single lines surrounded by double lines (left) and of a
region with double lines surrounded by single lines (right) in increasing
out-of-plane magnetic field, measured with an out-of-plane spin-sensitive Cr
bulk tip. When the external field increases, the spin spirals in the double
line regions transform into distorted skyrmions whereas those in the areas with
single lines become inhomogeneous and can be described as an assembly of
independent \SI{360}{\degree} domain walls. When the width of these walls is
similar to the width of the skyrmions in the adjacent areas, they couple to
them. 
\emph{Measurement~parameters:}
Left:~${U = \SI{-700}{\milli\volt}}$, ${I = \SI{1}{\nano\ampere}}$,
${T = \SI{8}{\kelvin}}$;
Right:~${U = \SI{-500}{\milli\volt}}$, ${I = \SI{1}{\nano\ampere}}$,
${T = \SI{4}{\kelvin}}$.
}
\end{figure}

In external out-of-plane magnetic fields between \SI{1}{\tesla} and
\SI{3}{\tesla}, the spin spirals in the double line regions split up into
distorted magnetic skyrmions~\cite{hsu_electric_2016}, as illustrated in
Fig.~\ref{adj_regions}. In contrast, no skyrmion is created on the single
line regions. There, the spirals become inhomogeneous and behave like an
assembly of independent \SI{360}{\degree} domain walls which disappear one by
one when the field is increased. Adding a Zeeman term to the model proposed
before (equation~(\ref{energy_density})) does not allow to understand this
difference. For a isotropic film without effective anisotropy, there is a
stable skyrmionic phase for any couple $(A,D)$ under the appropriate external
magnetic field. The crucial point for the triple layer Fe film on Ir(111) are
the dislocation lines which break the rotational symmetry. Thus one cannot
expect the existence of the typical circular shaped
skyrmions~\cite{hagemeister_pattern_2016}  as observed, for example, in the
PdFe bilayer on Ir(111)~\cite{romming_writing_2013} and a priori, even the
presence of skyrmions is not obvious. Nevertheless, distorted skyrmions
appear in the double line regions. Their bean-like shape is induced by the
local arrangement of the Fe atoms similarly to the zigzag shape of the spin
spiral wavefront. They are always located on top of three dislocation lines,
two identical ones at the ends and a third different one in the center. This
particular preferred position results from the local variation of the magnetic
interaction within the layer. The skyrmions are hence pinned to the lines and
naturally aligned on ``tracks" (see Fig.~\ref{adj_regions} and
Fig.~\ref{hysteresis}(g)). The atom arrangement is different for the single
lines and it appears that the skyrmion pinning effect is absent and
only \SI{360}{\degree} domain walls are observed in magnetic field.

\section{Metastability and transition fields}

\begin{figure}
	\includegraphics[scale=1]{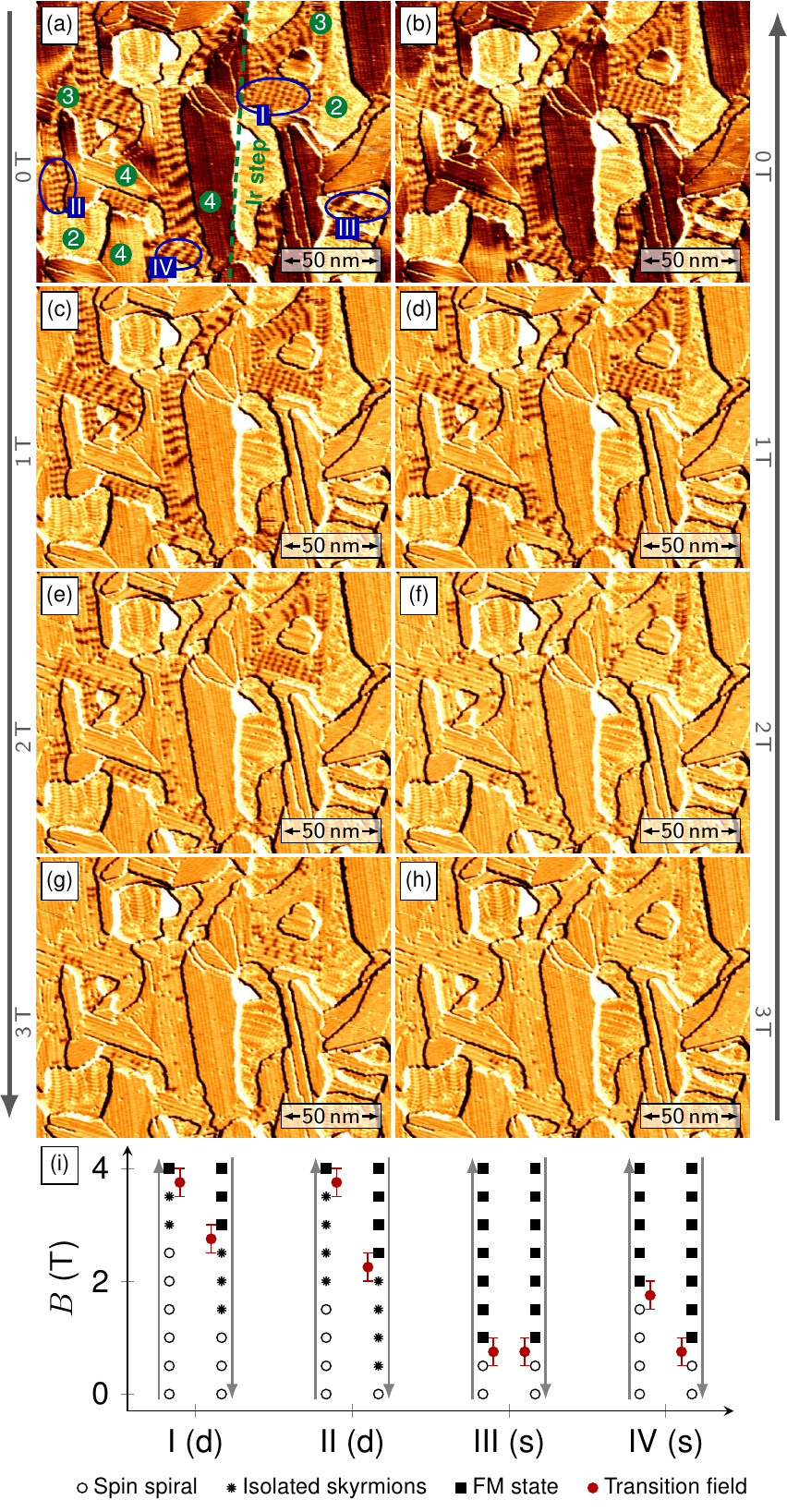}
	\caption{\label{hysteresis} (a)-(h) Spin-polarized differential tunneling
conductance maps of an ultrathin Fe film on Ir(111). The numbers in green
circles in (a) indicate the local Fe coverage. An external out-of-plane
magnetic field was applied, increased in steps of \SI{0.5}{\tesla} up to
\SI{4}{\tesla} (at this field the sample is completely polarized, i.e.
ferromagnetic) and then decreased again to zero. Comparison between scans taken
at the same field value shows a hysteretic behavior for the triple layer Fe,
both in the regions with the double and single lines.
(i) Plot of the magnetic state during the field sweep of the four areas marked
in blue in (a). Areas I and II are double line regions, III and IV single line
regions.
\emph{Measurement~parameters:} ${U=\SI{-500}{\milli\volt}}, {I=\SI{1}{\nano\ampere}},
{T=\SI{4}{\kelvin}}$, out-of-plane spin-sensitive Cr bulk tip. 
}
\end{figure}

A more quantitative investigation of the influence of an external magnetic 
field is shown in Fig.~\ref{hysteresis}. For this measurement series, the
magnetic field was increased up to \SI{4}{\tesla} in \SI{0.5}{\tesla} steps.
At \SI{4}{\tesla}, the sample has fully reached the ferromagnetic state. Then,
the magnetic field was decreased again in steps to zero. Every region in the
film behaves differently because of its structure and its interactions with
adjacent areas (see Fig.~\ref{adj_regions}). Comparisons between pictures taken
at the same field value during the up-sweep and the down-sweep reveal, however,
that almost all the areas show hysteresis. The skyrmions and domain walls 
vanish at a much higher field than the one needed to produce them again.
This indicates that the magnetic states are metastable.
	
The transition fields to the ferromagnetic (FM) state are obtained from field
dependent measurements using the procedure described in
Fig.~\ref{hysteresis}(i): the evolution of the magnetic states for different
regions is indicated for an up- and a down-sweep of the magnetic field. The
transition field value corresponds to the middle of the corresponding step and
the error bar to the step height. 

\begin{figure}[H]
\centering
	\includegraphics[scale=1]{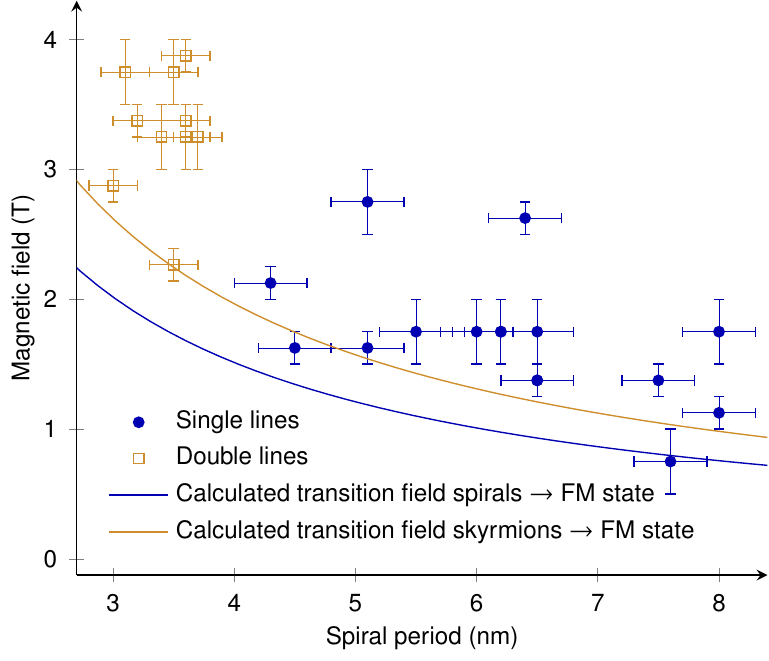}
	\caption{\label{critical_field} Effect of the spiral period on the magnetic
fields needed to reach the ferromagnetic state during an increasing field
sweep. The correlation between the spiral period and spacing of the dislocation
lines in the Fe layer shown in Fig.~\ref{plot_mag_strain} leads also to a
dependence of the transition field on the strain relief. A trend that higher
external fields are required to destroy non-collinear spin structures with
smaller spatial period appears. The calculated transition fields were obtained
from the phase diagram detailed by Bogdanov and
Hubert~\protect\cite{bogdanov_thermodynamically_1994} with the parameters from
Fig.~\ref{plot_mag_strain}: ${D = \SI{2.8}{\milli\joule\per\square\meter}}$ and
${K_\mathrm{eff} = 0}$. The magnetic moment used is ${\mu = 2.7 \mu_\mathrm{B}}$
per Fe atom~\cite{heinze_spontaneous_2011}. Because of the metastability of the
magnetic states revealed by the hysteresis, the experimental field values are
upper estimates of the transition fields. The SP-STM measurements were
performed at \SI{4}{\kelvin} on three different samples and the external
magnetic field was increased in steps of \SI{0.5}{\tesla}. The error bars
correspond to the height of the steps as indicated in Fig.~\ref{hysteresis}(i).
}
\end{figure}

Results for several samples are gathered in Fig.~\ref{critical_field}, where
they are correlated with the spiral period in the absence of external magnetic
field. Only sweeps with increasing field are considered in this figure, the
field values are thus upper estimates due to the hysteresis effect. The
transition field $B_\mathrm{t}$ is decreasing when the spiral period
increases. The trend that higher external fields are required to destroy
non-collinear spin structures with smaller spatial period is consistent which
the observation that the spin spiral in the double layer Fe does not change in
magnetic fields up to \SI{9}{\tesla}~\cite{hsu_guiding_2016}.

Although the simplified model~(\ref{energy_density}) does not help to
understand the absence of a skyrmionic phase in the single line regions, it
reproduces the decrease of the transition field for larger magnetic
structures once the Zeeman term 
${\mathcal{E}_\mathrm{z} = -M_\mathrm{s}B\mathrm{m}_z}$
is included. In the zero effective anisotropy case, the transition field can be
obtained from the phase diagram provided by Bogdanov and Hubert
\cite{bogdanov_thermodynamically_1994}. There is a threshold value for the
reduced field parameter $h$ such as:
\begin{equation}
  	B_\mathrm{t} = \frac{D^2 h_\mathrm{t}}{A M_\mathrm{s}} 
  	             = 4\pi\frac{Dh_\mathrm{t}}{\lambda M_\mathrm{s}}
\end{equation}       
The saturation magnetization  
${M_\mathrm{s} = \SI{1.77}{\mega\ampere\per\meter}}$ is estimated from the
magnetic moment of 2.7 $\mu_\mathrm{B}$ per atom in the monolayer Fe~\cite{heinze_spontaneous_2011}. We did not consider here a potential
variation of $M_\mathrm{s}$ with the strain relief. The threshold value
$h_\mathrm{t}$ is different for the transition from spirals to the FM state and
from skyrmions to the FM state:
\begin{align}
 	h^\mathrm{spiral}_\mathrm{t} & = 0.308 \\
 	h^\mathrm{skyrmion}_\mathrm{t} & = 0.401 
\end{align}

The transition fields are plotted as plain lines in
Fig.~\ref{critical_field}, assuming again that
${D = \SI{2.8}{\milli\joule\per\square\meter}}$. They are defined as the fields
at which the energy of the FM state is equal to the one of the spiral or
skyrmion state, respectively. As expected, the experimental values are larger
than the computed ones because of the metastability of the spiral and skyrmion
states in increasing magnetic field. Remarkably, the smallest field values are
almost on the theoretical curve and none of them below, which indicates a
rather good agreement between the model and the actual system. This behavior
of the transition fields supports our assumption that the effective exchange
coupling is affected by the strain relief and is responsible for the observed
variation of the magnetic properties of the Fe film.

\section{Conclusion}

Exploiting the strain relief in epitaxial ultrathin films is an effective way
to control precisely and locally their magnetic state. Both the typical size of
the spin structures and the transition fields could be tuned. Moreover, the
actual uniaxial structure and pinning properties of ultrathin films
exhibiting dislocation lines may allow to stabilize a skyrmion state as well as
to confine skyrmions on well-defined tracks on the order of a few
nanometers. Combined with the possibility to write and delete the
skyrmions by a local electric field~\cite{hsu_electric_2016}, this could be of
great interest in view of future racetrack-based spintronic devices~\cite{parkin_magnetic_2008, fert_skyrmions_2013, wiesendanger_nanoscale_2016}.

\begin{acknowledgments}

Financial support by the European Union via the Horizon 2020 research and
innovation programme under grant agreement No. 665095, by the Deutsche
Forschungsgemeinschaft via SFB668-A8, and by the Hamburgische Stiftung
f\"{u}r Wissenschaften, Entwicklung und Kultur Helmut und Hannelore Greve is
gratefully acknowledged.

\end{acknowledgments}

\end{document}